\documentstyle[12pt,epsfig]{article}
\begin{document}
%
% Here you can put your personal macros at will. E.g.:
\def    \be             {\begin{eqnarray}}
\def    \ee             {\end{eqnarray}}
%
% Your contribution should appear as a Section:

\begin{flushright}
Freiburg--THEP 95/20\\
November 1995
\end{flushright}

\vspace{1.5cm}

\begin {center}
{\large \bf  The hidden Higgs model at LEP2
 \footnote{Contribution to the Workshop on Physics at LEP2.}}
\vskip 1.cm
{\bf T. Binoth }\\ \medskip Albert--Ludwigs--Universit\"at Freiburg\\
Fakult\"at f\"ur Physik \\ Hermann--Herder--Strasse 3\\
79104 Freiburg i. Br.\\
\bigskip
{\bf J. J. van der Bij }\\ \medskip Albert--Ludwigs--Universit\"at Freiburg\\
Fakult\"at f\"ur Physik \\ Hermann--Herder--Strasse 3\\
79104 Freiburg i. Br.\\
and \\
CERN, Theory Division \\
CH--1211 Gen\`{e}ve 23\\
Switzerland
\end{center}

\bigskip
\begin{abstract}
The influence of massless scalar singlets on the Higgs signal
            at LEP2 is discussed. It is shown that for strong interactions
            between the Higgs boson and the singlet fields, detection of the
Higgs
            signal can become impossible.
\end{abstract}

\newpage
\section{Introduction}
The radiative corrections at LEP1 depend only logarithmically on the Higgs
mass,
and the measurements, although very precise, are not sufficient
to determine the structure of the Higgs sector. It is therefore necessary
to keep an open mind to the possibility that the Higgs sector is
more complicated than in the Standard Model.
Beyond the Standard Model various extensions have been suggested.
One of the possibilities is supersymmetry which is discussed
in a previous section. Another possibility is strong
interactions in the form of technicolor, which at
least in its simplest form is ruled out by the LEP1 data.
Strong interactions in the Standard Model itself imply a heavy Higgs boson
and can presumably be studied at the LHC.

However
the idea of strong interactions is more general. In particular
it is possible that strong interactions are present in the
singlet sector of the theory. In general the  choice of representations
in a gauge theory is arbitrary and presumably a clue to a deeper
underlying theory. Singlets do not have
quantum numbers under the gauge group of the Standard Model.
They therefore do not feel the strong or electro--weak forces,
but they can couple to the Higgs particle. As a consequence
radiative corrections to weak processes are not sensitive to the
presence of singlets in the theory, because no Feynman graphs containing
singlets  appear
at the one--loop level. Because effects at the two--loop level
are below the experimental precision,
the presence of a singlet sector is not ruled out by any
of the LEP1 precision data.

It is therefore not unreasonable to
assume that there exists a hidden sector, that affects Higgs
physics only. Such an extension of the Standard Model involving
singlet fields preserves the essential simplicity of the model,
while at the same time acting as a realistic model for non--standard Higgs
properties. Here we will study the coupling
of a Higgs boson to an O(N) symmetric set of scalars, which
is one of  the simplest possibilities introducing only a few extra
parameters in the theory. The effect of the extra scalars is practically
the presence of a large invisible decay width of the Higgs particle.
When the coupling is large enough the Higgs resonance can become
wide even for a light Higgs boson. This has led to the conclusion that
this Higgs particle becomes undetectable at the LHC \cite{vladimir}.
As one can measure missing energy
more precisely at $e^+e^-$--colliders than at a hadron machine,
LEP2 can give important constraints
on the parameters of the model. However it is clear that there
will be a range of parameters, where this Higgs boson can be seen neither
at LEP nor at the LHC. In the next section we will introduce the
model together with presently known constraints and in the last section
we will discuss exclusion limits at LEP2.
\section{The model}
The Higgs sector of the model is described by the following Lagrangian,
\be
\label{definition}
 {\cal L}  =
 - \partial_{\mu}\phi^+ \partial^{\mu}\phi -\lambda (\phi^+\phi - v^2/2)^2
   - 1/2\,\partial_{\mu} \vec\varphi \partial^{\mu}\vec\varphi
     -1/2 \, m^2 \,\vec\varphi^2 - \kappa/(8N) \, (\vec\varphi^2 )^2
    -\omega/(2\sqrt{N})\, \, \vec\varphi^2 \,\phi^+\phi \nonumber
\ee
where $\phi$ is the normal Higgs doublet and the vector $\vec\varphi$
is an N--component real vector of scalar fields, which we call phions.
Couplings to fermions and vector bosons are the same as in the Standard Model.
The ordinary
Higgs field acquires the vacuum expectation value $v/\sqrt{2}$.
We assume that the $\vec\varphi$--field acquires no vacuum expectation
value, which can be assured by taking $\omega$ positive. After the spontaneous
symmetry breaking one is left with the ordinary Higgs boson,
coupled to the phions in which it decays. Also the phions
receive an induced mass from the spontaneous symmetry breaking.
The factor N is taken
to be large, so that the model can be analysed in the $1/N$ expansion.
By taking this limit the phion mass stays small, but because there
are many phions the decay width of the Higgs boson can become large.
Therefore the main effect of the presence of the phions is to give
a large invisible decay rate to the Higgs boson. The
invisible decay width is given by
\be \Gamma_H =\frac {\omega^2 v^2}{32 \pi M_H} \quad .\nonumber \ee
The Higgs width is compared with the width in the Standard Model for various
choices
of the coupling $\omega$ in Fig.~\ref{width}.
The model is different
from Majoron models \cite{valle}, since the width is not necessarily small.
The model is similar to the technicolor--like model of \cite{chivukula}.
\begin{figure}[h]
\vspace{0.1cm}
%\centerline{\epsfig{figure=width.eps,height=8.5cm,angle=0}}
\caption{\it Higgs width in comparison with the Standard Model.}
\label{width}
\end{figure}

Consistency of the model requires two conditions.
One condition is the absence of a Landau pole below a certain scale
$\Lambda$. The other follows from the stability of the vacuum up to a certain
scale. An example of such limits is given in Fig.~\ref{stability},
where $\kappa=0$ was taken at the scale $2m_Z$, which allows for
the widest range. For the model to be valid beyond  a scale
$\Lambda$ one should be below the indicated upper lines in the figure,
as otherwise there would appear a Landau pole before this scale.
One should be to the right of the indicated lower lines
to have stability of the vacuum.
\begin{figure}[h]
\vspace{0.1cm}
%\centerline{\epsfig{figure=stability.eps,height=8.5cm,angle=0}}
\caption{\it Theoretical limits on the parameters of the model
in the $\omega$ vs. $M_H$ plane. Beyond a scale $\Lambda$, the physical region
is below the indicated upper lines and to the right of the lower
lines.}
\label{stability}
\end{figure}

For the search for the Higgs boson there are basically
two channels, one is the standard decay, which is reduced in branching
ratio due to the decay into phions.
The other is the invisible decay, which rapidly becomes dominant,
eventually making the Higgs resonance wide (see Fig.~\ref{width}).
In order to give the bounds we
neglect the coupling $\kappa$ as this is a small effect. We
also neglect the phion mass. For other values of the phion mass
the bounds can be found by rescaling the decay widths
with the appropriate phase space factor.
The present bounds, coming from LEP1 invisible search,
are included as a dashed curve in Fig.~\ref{exclusion} below.

\section{LEP2 bounds}
In the case of LEP2 the limits on the Higgs mass  and couplings in the present
model
come essentially from the invisible decay, as the branching ratio
into $\bar bb$ quarks drops rapidly with increasing $\varphi$--Higgs
coupling. To define the signal we look at events around the maximum of the
Higgs
resonance, with an invariant mass $m_H \pm \Delta$, with
$\Delta=5$ GeV, which corresponds to a typical mass resolution.
Exclusion limits are determined by Poisson statistics
as defined in the Interim Report \cite{LEP2WG}. The results are given by the
full lines
in Fig.~\ref{exclusion}. One notices the somewhat reduced sensitivity
for a Higgs mass near the Z boson mass and  a looser bound for small Higgs
masses,
because there the effect of widening the resonance is bigger.
The small $\omega$ region is covered by visible search.
There is
a somewhat better limit on the Higgs mass for moderate $\omega$ in
comparison with the $\omega=0$ case; this is due to events from
the extended tail of the Higgs boson resulting from the increased width.

We conclude from the analysis that LEP2 can put significant
limits on the parameter space of the model. However there is a range
where the Higgs boson will not be discovered, even if it does exist in this
mass range.
This holds also true, when one considers the search at the LHC.
Assuming  moderate to large values of $\omega$, i.e. in the already
difficult intermediate mass range, it is unlikely that sufficient signal events
are left at the LHC. In that case the only information can come directly
from the NLC or indirectly from higher precision experiments at LEP1.
\begin{figure}[h]
\vspace{0.1cm}
%\centerline{\epsfig{figure=exclusion.eps,height=8.5cm,angle=0}}
\caption{\it Exclusion limits at LEP2 (full lines), and LEP1 (dashed). The
region
where $\omega$ is small is covered by the search for visible Higgs decay.}
\label{exclusion}
\end{figure}
\end{document}